\documentclass[a4paper,10pt]{article}
\usepackage{enumerate}
\usepackage{color}
\usepackage[utf8]{inputenc} 
\usepackage[english]{babel}
\usepackage[T1]{fontenc}
\usepackage{graphicx}
\usepackage{amsfonts,amssymb,amsmath,latexsym,amsthm}
\usepackage{textcomp}
\usepackage[pdftex]{hyperref}
\usepackage{geometry}
\DeclareMathOperator{\sech}{sech}
\DeclareMathOperator{\sign}{sgn}

\title{Interactions of solitons with an external force field: Exploring the Schamel equation framework}
\author{Marcelo V. Flamarion$^{1}$ and Efim Pelinovsky$^{2,3}$}
\date{}

\begin{document}
\maketitle
\begin{center}
{\footnotesize $^1$Unidade Acad{\^ e}mica do Cabo de Santo Agostinho, \\
UFRPE/Rural Federal University of Pernambuco, BR 101 Sul, Cabo de Santo Agostinho-PE, Brazil,  54503-900 \\
marcelo.flamarion@ufrpe.br }

\vspace{0.3cm}
{\footnotesize $^{2}$Institute of Applied Physics, 46 Uljanov Str., Nizhny Novgorod 603155, Russia. \\
 $^{3}$National Research University--Higher School of Economics, Moscow, Russia. }


\end{center}


\begin{abstract} 
This study aims to investigate the interactions of solitons with an external force within the framework of the Schamel equation, both asymptotically and numerically. By utilizing asymptotic expansions, we demonstrate that the soliton interaction can be approximated by a dynamical system that involves the soliton amplitude and its crest position. To solve the Schamel equation, we employ a pseudospectral method and compare the obtained results with those predicted by the asymptotic theory. Remarkably, our findings reveal a qualitatively agreement between the predictions and the numerical simulations at early times. Specifically, we classify the soliton interaction into three categories: (i) steady interaction occurs when the crest of the soliton and the crest of the external force are in phase, (ii) oscillatory behavior arises when the soliton's speed and the external force speed are close to resonance, causing the soliton to bounce back and forth near its initial position, and (iii) non-reversible motion occurs when the soliton moves away from its initial position without changing its direction.

	\end{abstract}

\section{Introduction}
The Schamel equation is  a mathematical model frequently used in the field of plasma physics to describe nonlinear wave phenomena. Specifically, it is utilized to study the dynamics of ion acoustic waves in plasmas \cite{Schamel:1972, Schamel:1973, Ali:2017, Chowdhury:2018, Mushtaq:2006, Williams:2014}. It was first derived by Schamel \cite{Schamel:1973}  by considering the interaction between ions and electrons in a plasma, taking into account the effects of collisions, plasma density, and temperature gradients. The Schamel equation is particularly relevant in investigating phenomena such as solitons, which are solitary, self-reinforcing waves that propagate in plasmas. Solitons in the context of the Schamel equation can exhibit interesting behaviors such as self-trapping and particle acceleration. External perturbations have been observed to have an effect in certain real-life physical scenarios \cite{Nozaki:1983, Williams:1997, Beiglbock:2000}. Moreover, it has been found that these perturbations can vary under different physical conditions. Recent studies have placed particular emphasis on investigating nonlinear traveling wave solutions  in the presence of such external  perturbations \cite{Saha:2015a, Saha:2015b}. Recently, Chowdhury et al. \cite{Chowdhury:2018} derived a Schamel-type equation in the presence of an external force. They investigated the interaction of a soliton with a time-dependent external periodic force, asymptotically and numerically. The primary distinction between the renowned Korteweg-de Vries equation and the Schamel equation lies in the representation of nonlinearity, wherein the modular term characterizes the latter. This crucial distinction renders the Schamel equation to be a nonintegrable equation and introduces mathematical challenges associated with the non-analyticity of the function. 

The study of soliton interactions with an external force has garnered significant interest and has been extensively investigated over the past decades. This external force typically represents a localized topography or a moving source  \cite{Baines}. Researchers have explored this problem within various frameworks, including the forced Korteweg-de Vries equation (fKdV) \cite{Ermakov, Lee, LeeWhang, Kim, COAM, Collisions, Capillary2, Malomed:1993, Grimshaw:1993, Wu1, Paul,  Grimshaw86}, the Euler equations \cite{COAM2}, the forced modified Korteweg-de Vries equation (fmKdV) \cite{Chaos:FP,Flamarion-Pelinovsky:2022a}, and the non-integrable forced Whitham equation \cite{Whitham1, Whitham2}.
One intriguing phenomenon observed during soliton interactions with external forces is the occurrence of trapped waves. Trapped waves are characterized as waves that undergo a back-and-forth bouncing motion at the external force, remaining trapped for extended periods of time. This phenomenon emerges when the speed of the solitary wave and the speed of the external force are nearly in resonance.

An extensive asymptotic investigation of trapped waves in the context of the fKdV equation was conducted by Grimshaw and colleagues for a localized external force \cite{Grimshaw94, Grimshaw96}. Within the asymptotic framework, three distinct regimes were identified. Firstly, a soliton  can exhibit a bouncing behavior at the external force, remaining trapped for extended periods of time. Secondly, it can pass over the external force without reversing its direction. Finally, it may remain steady at the external force. Notably, these studies revealed a strong agreement between the asymptotic analysis and numerical predictions. Grimshaw and his team demonstrated that the asymptotic framework provided reliable and accurate results when compared with numerical simulations.


The objective of this article is to  examine the interaction between solitons and an externally imposed  force that moves with constant speed through the forced Schamel equation. The inclusion of an external force in the Schamel equation allows for a more comprehensive examination of the system response to external perturbations and provides insights into the interplay between the intrinsic dynamics of the plasma and external influences. This investigation is conducted through numerical and asymptotic approaches. The obtained asymptotic and numerical results are compared qualitatively and regimes of trapping conditions are discussed. 

This article is organized as follows. In section 2 we present the mathematical formulation of the problem. The results  are presented in section  3 and the conclusion is presented in section 4.
%
%
%

\section{The forced Schamel equation}
We study the interaction of solitons with an external force field by considering the Schamel equation in canonical form, including an external force $f(x)$ with a constant speed $\Delta$

\begin{equation}\label{Schamel1}
u_{t} +\sqrt{|u|}u_{x}+u_{xxx}=f_{x}\Big(x+\Delta t\Big).
\end{equation}

In this equation, $u(x,t)$ represents the wave field and $f(x)$ represents the external force. We aim to investigate how solitons interact with this external force field. To this end, it is convenient to rewrite equation (\ref{Schamel1}) in the external force moving frame.  Therefore, we write the traveling variables $$x'=x+\Delta t, \;\ t'=t.$$
In the new coordinate system, dropping the primes equation (\ref{Schamel1})  can be expressed as
\begin{equation}\label{Schamel2}
u_{t} +\Delta u_{x}+\sqrt{|u|}u_{x}+u_{xxx}=f_{x}(x).
\end{equation}
Before proceeding, it is useful to note that the forced Schamel equation is a Hamiltonian equation, with Hamiltonian
\begin{equation}\label{Hamiltonian}
\mathcal{H} = \int_{-\infty}^{+\infty}\Big[\frac{\Delta u^{2}}{2}-\frac{1}{2}u_{x}^{2}+\frac{4}{15}\sign(u)|u|^{5/2}-uf(x)\Big] dx.
\end{equation}
In terms of the functional $\mathcal{H}$, equation (\ref{Schamel2}) can be written in the Hamiltonian form
\begin{equation*}
u_{t} = \frac{\partial}{\partial x}\Big[\frac{\delta\mathcal{H}}{\delta u}\Big],
\end{equation*}
where
\begin{equation*}
\frac{\delta\mathcal{H}}{\delta u}=\Delta u+u_{xx}+\frac{2}{3}\sign(u)|u|^{3/2}- f(x).
\end{equation*}
Further, since $\mathcal{H}$ contains no explicit time dependence, the Hamiltonian $\mathcal{H}$ is an invariant. Also equation (\ref{Schamel2}) possesses the Casimir invariant which is usually called as the mass invariant given by the formula
\begin{equation}\label{mass}
M(t) = \int_{-\infty}^{+\infty}u(x,t) dx.
\end{equation}
The formulas (\ref{Hamiltonian})-(\ref{mass}) serve a valuable purpose, particularly in assessing the precision and reliability of numerical methodologies employed for solving the Schamel equation (\ref{Schamel2}).

When there is no external force present, the Schamel equation (\ref{Schamel2}) admits solitons as solutions. These solitons can be described by the following expressions
\begin{equation}\label{solitary}
u(x,t)=a\sech^{4}\Big(k(x-ct)\Big), \mbox{  where }  \; c =\Delta+\frac{8\sqrt{|a|}}{15} \mbox{ and } k = \sqrt{\frac{c}{16}}.
\end{equation}

Here, $a$ represents the soliton amplitude, which can also be negative, $c$ represents the soliton speed, and $k$ characterizes the soliton wavenumber.

In our investigation, we consider the external force to be of the form
\begin{equation}\label{externalforce}
f(x) = b \exp\Big(-\frac{x^{2}}{w^2}\Big).
\end{equation}
In this expression, $w$ represents the width of the external force and $b$ represents its amplitude which can also be negative. Notice that if $u(x,t)$ is a solution of equation (\ref{Schamel2}) for an external force term $f(x)$, then $-u(x,t)$  is a solution for the same problem with external force term $-f(x)$. Therefore, we restrict ourselves to considering only the case $b>0$ and similar results are valid for $b<0$.

\section{Results}
\subsection{Asymptotic results}
In this section, we aim to derive the governing equations of the  interaction between solitons and an external force, assuming that the force has small amplitude. To this end, we introduce a small positive parameter, denoted as $\epsilon$, and replace the external force $f$ with $\epsilon f$ in equation (\ref{Schamel2}). Additionally, we make the assumption that the wave field closely resembles a soliton, with slowly varying parameters over time \cite{Grimshaw94, Grimshaw2002, Pelinovsky:2002}. The soliton can be mathematically described using the following formulas
\begin{equation}\label{solitary}
u(\mathbf{\Phi},T)=a(T)\sech^{4}(k(T)\mathbf{\Phi}), \mbox{  where }  \;\ \mathbf{\Phi}=x-X(T) \mbox{ and } X(T) =x_0+\frac{1}{\epsilon}\int_{0}^{T} q(T)dT,
\end{equation}
where $x_0$ is the initial position of the soliton, the functions $a$ and $q$ are determined from the interaction between the wave field and the external field. Here, we  formally introduce the ``slow time" by considering the new variable $T=\epsilon t$. We seek for a solution in the form of the asymptotic expansion 
\begin{align} \label{Asymptotic}
\begin{split}
& u(\mathbf{\Phi},T)=u_{0}+\epsilon u_1+\epsilon^{2} u_2+\cdots , \\
& q(T) = q_0 + \epsilon q_1 + \epsilon^{2} q_2+\cdots. \\
\end{split}
\end{align}
At the lowest order of the perturbation theory, the solutions $u_0$ and $q_0$  are defined as in equation (\ref{solitary}).

The first-order momentum balance is given by the expression
\begin{equation}\label{momentum}
\frac{1}{2}\frac{d}{dT}\int_{-\infty}^{\infty} u_{0}^{2}(\mathbf{\Phi})d\mathbf{\Phi} = \epsilon\int_{-\infty}^{\infty}u_{0}(\mathbf{\Phi})\frac{df}{d\mathbf{\Phi}}(\mathbf{\Phi}+X)d\mathbf{\Phi}. 
\end{equation}
Substituting the formulas (\ref{solitary}) into equation (\ref{momentum}) we obtain the dynamical system

\begin{align} \label{DS0}
\begin{split}
& \frac{da}{dt} = \frac{35}{32}k\int_{-\infty}^{\infty}\sech^{4}(k\mathbf{\Phi})\frac{df}{d\mathbf{\Phi}}(\mathbf{\Phi}+X)d\mathbf{\Phi}, \\
& \frac{dX}{dt}=\Delta+\frac{8}{15}\sqrt{|a|}.
\end{split}
\end{align}

When the external force is significantly broader than the soliton length, it is possible to approximate the soliton as a delta-function. As a result, the dynamical system governing the amplitude and position of the soliton crest can be simplified to the following form
\begin{align} \label{DS}
\begin{split}
& \frac{da}{dt}=\frac{35}{32}\frac{df}{dX}(X), \\
& \frac{dX}{dt} = \Delta+\frac{8}{15}\sqrt{|a|}.
\end{split}
\end{align}

Solutions of the dynamical system (\ref{DS}) have  equilibrium points located at the maxima and minima of the external force only if $\Delta$ is negative.  The amplitude of the resonance soliton ($a_0$) and the position of its crest $(X_0)$ are
\begin{equation} \label{equilibrium}
a_{0}^{\pm}=\pm\Big(\frac{15\Delta}{8}\Big)^2 \mbox{ and } X_{0}=0 \mbox{ for } \Delta<0.
\end{equation}
The equilibrium position is classified as a center when the disturbance and the soliton have the same polarity, while it is considered a saddle when they have opposite polarities. Centers represent steady solitons  and saddles represent solitons that move away from the external field without reversing their direction, and their interaction ceases once the waves pass over the external force.

The solutions of system (\ref{DS}) can be represented by streamlines, which are the level curves of the stream function
\begin{equation}\label{streamfunction}
\mathbf{\Psi}(X,a) = -\frac{35}{32}f(X)+\Delta a+\sign(a)\frac{16}{45}|a|^{3/2}.
\end{equation}

In order to analyse the phase portrait of system (\ref{DS})  for the external force (\ref{externalforce}) we rescale the variables as follows: the coordinate $X$ is rescaled with respect to $w$, and the amplitude $a$ with respect to $(35b/32)^{2/3}$, where $b>0$ is the amplitude of the external forcing. This yields the parameter 
\begin{equation}
\widetilde{\Delta}= \frac{\Delta}{((35/32)b)^{5/3}}.
\end{equation}
In this new coordinates system, the  streamfunction reads
\begin{equation}\label{streamfunction}
\mathbf{\Psi}(X,a) = -e^{-{X^2}}+\tilde{\Delta}a+\sign(a)\frac{16}{45}|a|^{3/2}.
\end{equation}

\begin{figure}[!h]
	\centering	
	\includegraphics[scale =1]{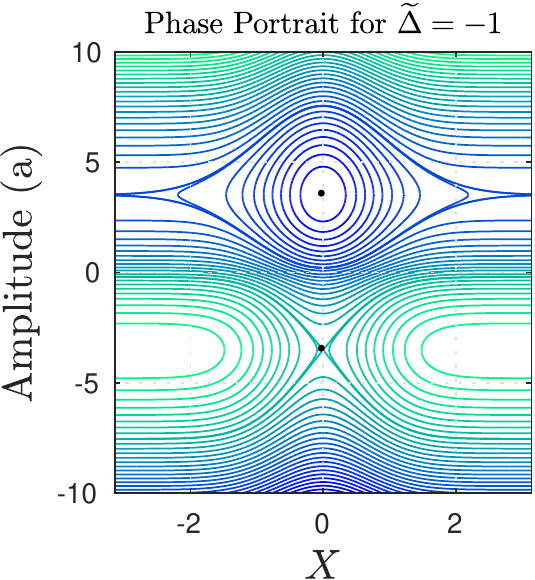}
 	\includegraphics[scale =1]{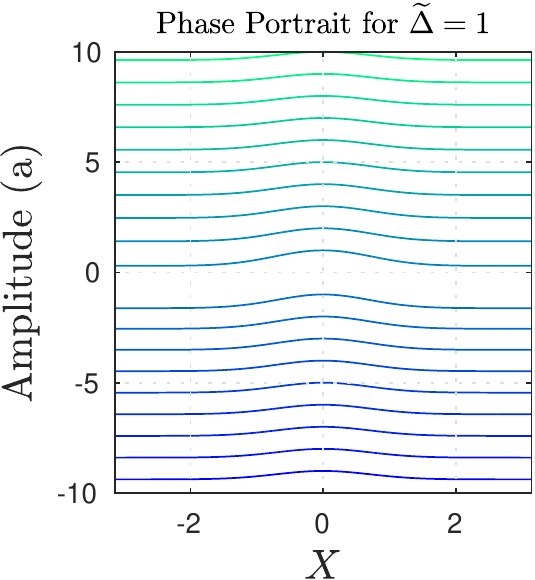}
	\caption{Phase portraits for the dynamical system (\ref{DS}). Circles correspond to equilibrium points.}
	\label{Fig1}
\end{figure}

Figure \ref{Fig1} depicts typical phase portraits of system (\ref{DS}). It is important to note that a closed orbit represents a soliton that is trapped without radiation, resulting from its interaction with the external force. On the other hand, a non-closed orbit signifies a soliton that propagates continuously without reversing its direction. It is important to point out that in Figure \ref{Fig1}, each equilibrium point aligns with a crest of the external force.

It is worth mentioning that the asymptotic method is not applicable in the vicinity of zero amplitude. This is because the soliton width increases without restriction, preventing it from changing slowly. Furthermore, according to the asymptotic theory, a soliton of one polarity cannot change its polarity. As the soliton amplitude approaches zero, all terms in the Schamel equation have the same order, causing perturbation theory to break down. This observation is supported by equation (\ref{momentum}), which characterizes the conservation of polarity.

\subsection{Numerical results}

\begin{figure}[h!]
	\centering	
	\includegraphics[scale =1]{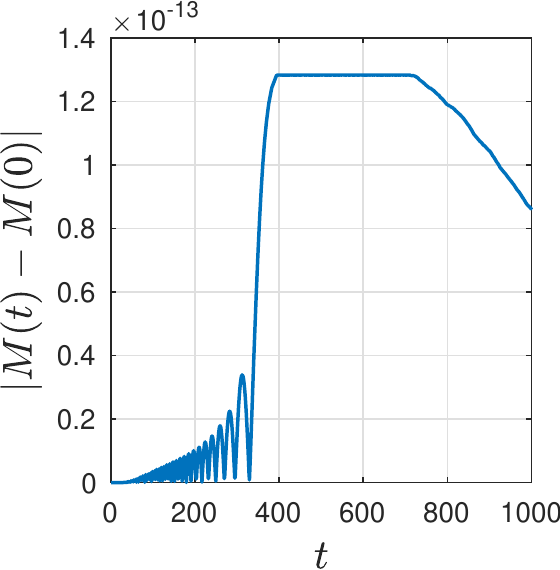}
	\includegraphics[scale =1]{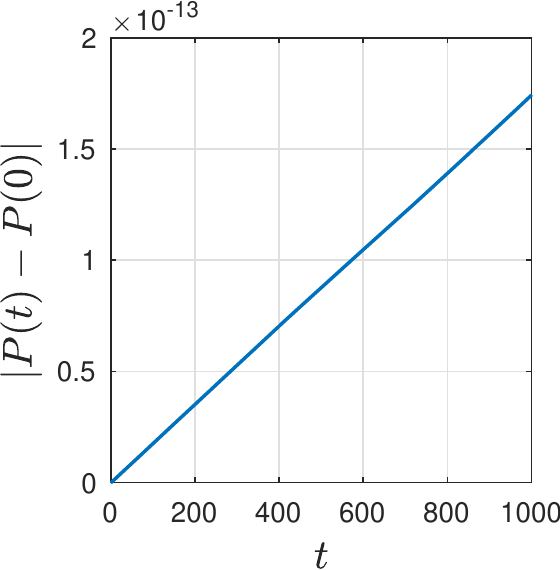}
	\caption{ The conserved quantities for the unforced problem with a soliton with amplitude $a=1$. On the left the mass conservation and on the right the momentum conservation. }
	\label{FigConserved}
\end{figure}

The numerical solution of equation (\ref{Schamel2}) is obtained by employing a Fourier pseudospectral method with an integrating factor. The equation is solved in a periodic computational domain $[-L, L]$ with a uniform grid containing $N$ points. This grid allows us to approximate the spatial derivatives accurately \cite{Trefethen:2000}.  To mitigate the influence of spatial periodicity, the computational domain is chosen to be sufficiently large. For the time evolution of the equation, we employ the classical fourth-order Runge-Kutta method with discrete time steps of size $\Delta t$. The external force is chosen as in equation (\ref{externalforce}) with $b=1$, $w=10$ and the parameter $\epsilon=0.01$. For a  complete study resolution of a similar numerical method, the readers are refer to the work of Flamarion et al. \cite{Marcelo-Paul-Andre}. Additionally, in the absence of the external force, we verified that the numerical method conserves mass and momentum. Figure \ref{FigConserved} displays the mass and momentum quantities throughout time for a initial soliton with amplitude $a=1$. 

In order to establish a comparison between the numerical solutions and the asymptotic predictions, we conduct an analysis to determine if the equilibrium points of the dynamical system (\ref{DS}) correspond to characteristic solutions of equation (\ref{Schamel2}). Our objective is to ascertain whether a point in the vicinity of a saddle point represents a soliton that undergoes unidirectional motion without reversing its direction, and whether a center point corresponds to a trapped soliton. This investigation allows us to validate the consistency between the numerical and asymptotic approaches in capturing the qualitative behavior of soliton dynamics.
\begin{figure}[h!]
	\centering	
	\includegraphics[scale =1]{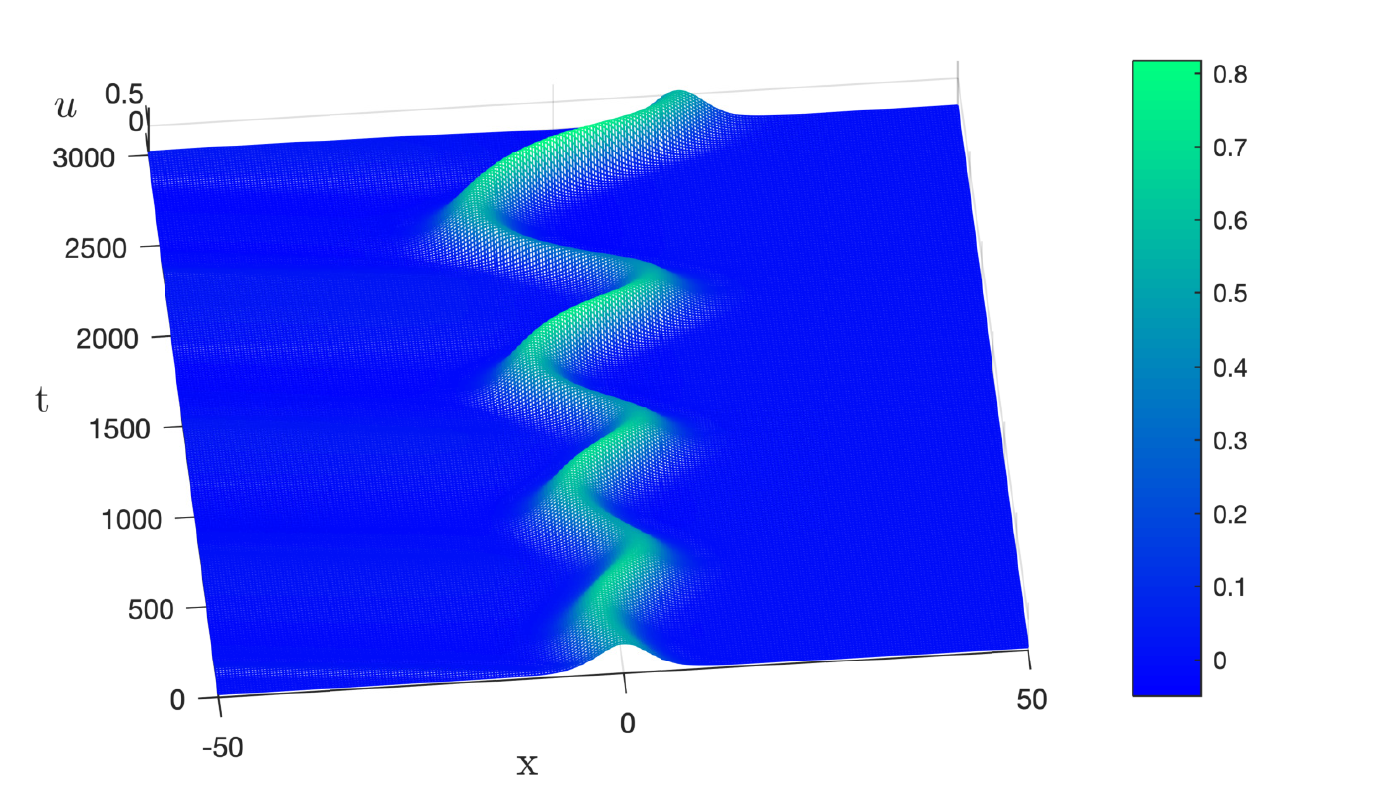}
	\includegraphics[scale =1]{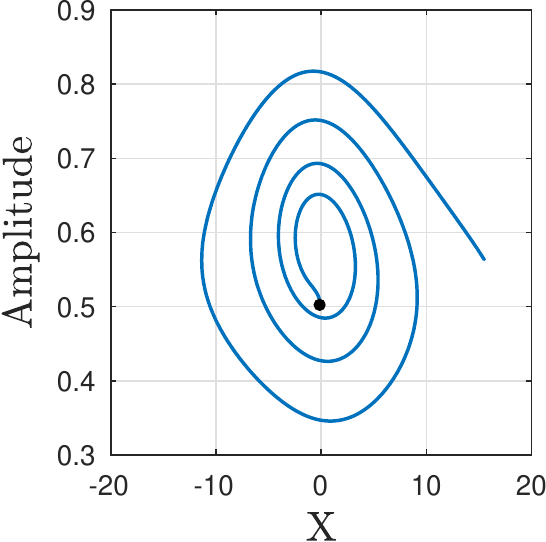}
	\includegraphics[scale =1]{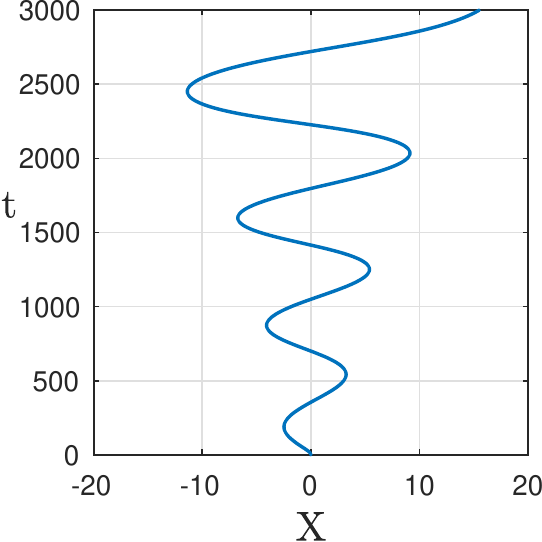}
	\caption{ Top: trapped soliton with the same polarity of the external force. Bottom (left): the space amplitude vs. crest position. Bottom (right): the crest position along the time. Parameters: $a=0.5$, $w=10$ and $\epsilon=0.01$ and  $\Delta=-\frac{8}{15}\sqrt{|a|}$.}
	\label{Fig2}
\end{figure}

\begin{figure}[h!]
	\centering	
	\includegraphics[scale =1]{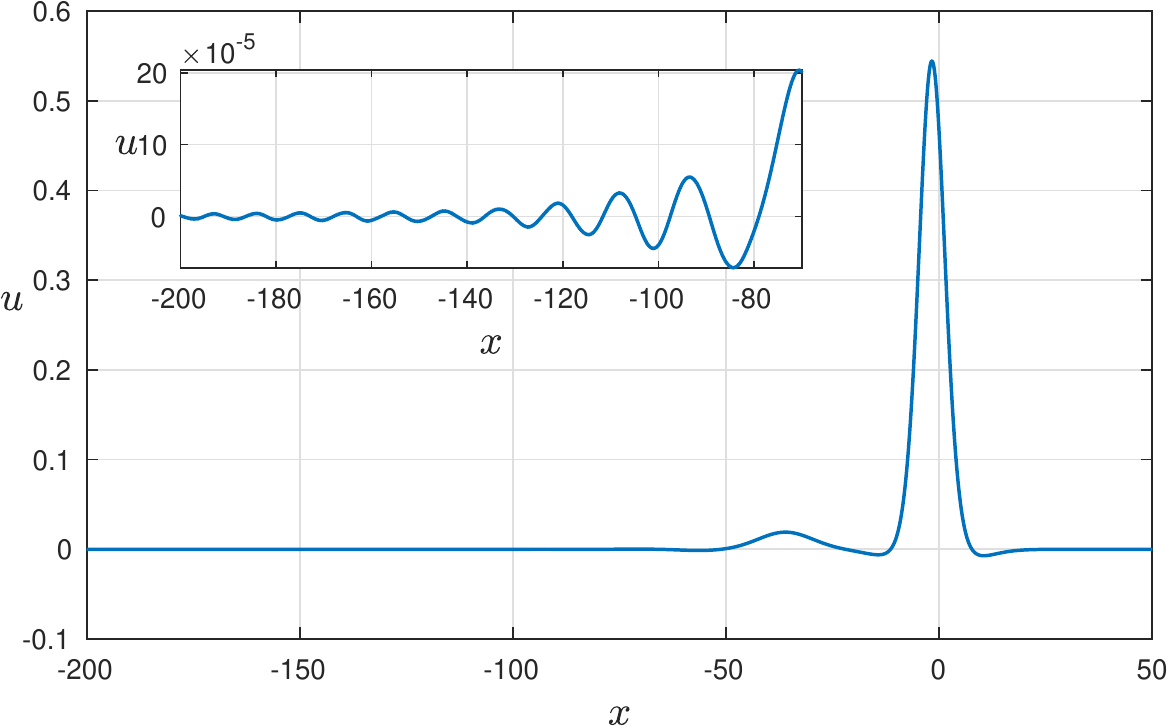}
	\caption{ A snapshot of Figure \ref{Fig2} (top) at time $t=100$. }
	\label{Fig3}
\end{figure}

We consider the interesting case when $\Delta$ is negative, as it allows for the trapping of solitons by the external field. The qualitative  agreement between the asymptotic theory predictions and the numerical solutions is noteworthy. It is worth mentioning that the asymptotic solution is truncated at the order of $\mathcal{O}(\epsilon)$, which sets the expected error when comparing the numerical and asymptotic solutions. Figure \ref{Fig2} illustrates a representative solution of equation (\ref{Schamel2}) for an initial soliton, with the amplitude and crest position chosen to be the center of the dynamical system (\ref{DS}). Notably, the crest of the soliton remains confined within a small region for extended periods of time, indicating nearly closed trajectories in amplitude versus crest position in space. However, the soliton moves away from the external force at large times which differs quantitatively from the asymptotic predictions. Moreover, in the position of crest versus time plot, the soliton behavior resembles a forced harmonic oscillator (unstable spiral). The dispersive tail produced during the interaction between the soliton and the external force is highlined in Figure \ref{Fig3} for a certain time.  This dispersive tail resembles linear Airy function which is a solution of linearized KdV. It is worth mentioning that similar results are observed for different initial choices of soliton amplitude and crest position, as long as they are close to the center point predicted by the asymptotic theory.

\begin{figure}[h!]
	\centering	
	\includegraphics[scale =1]{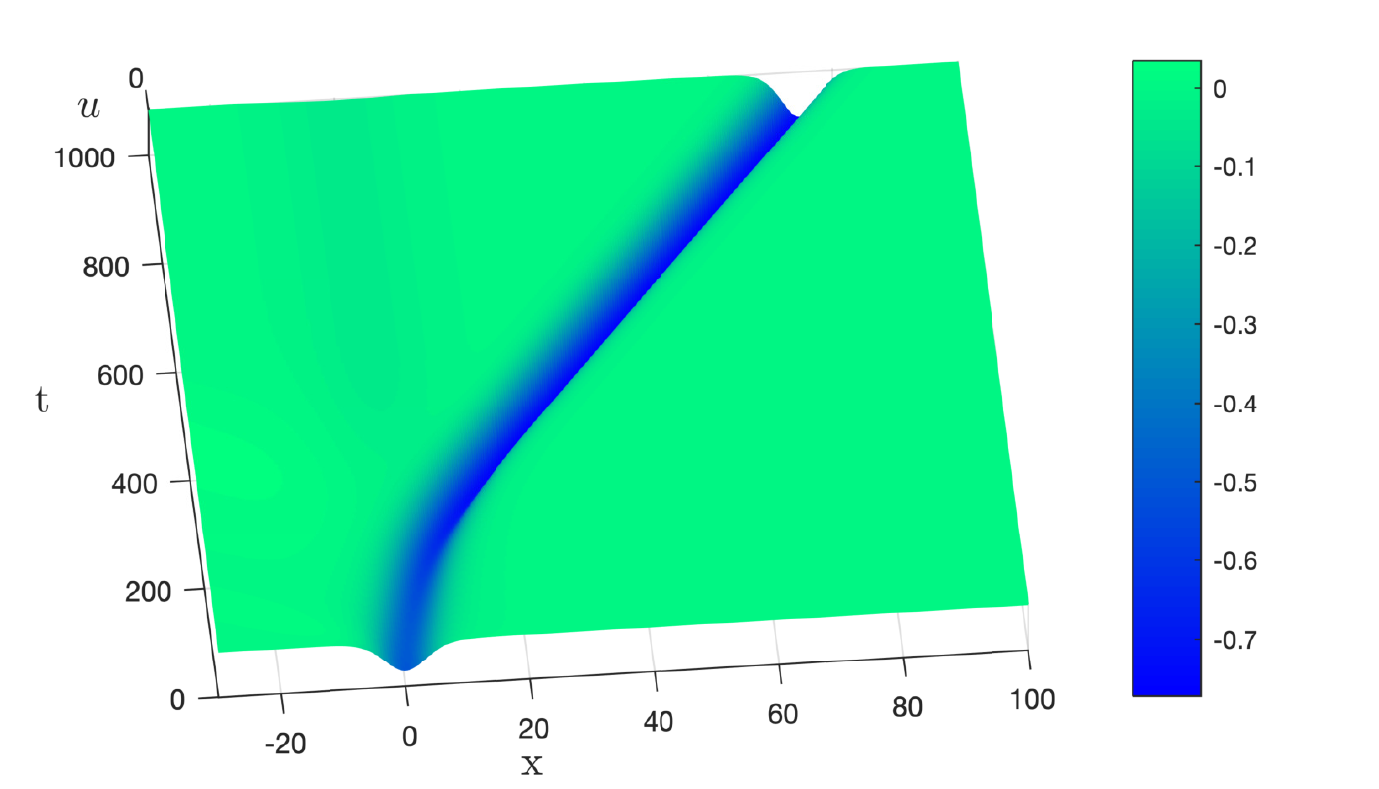}
	\includegraphics[scale =1]{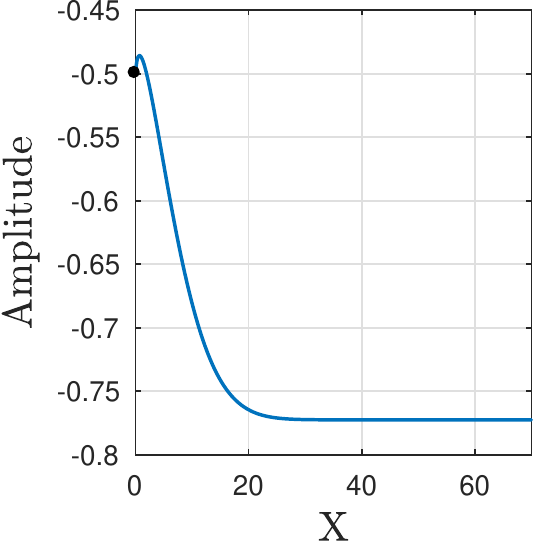}
	\includegraphics[scale =1]{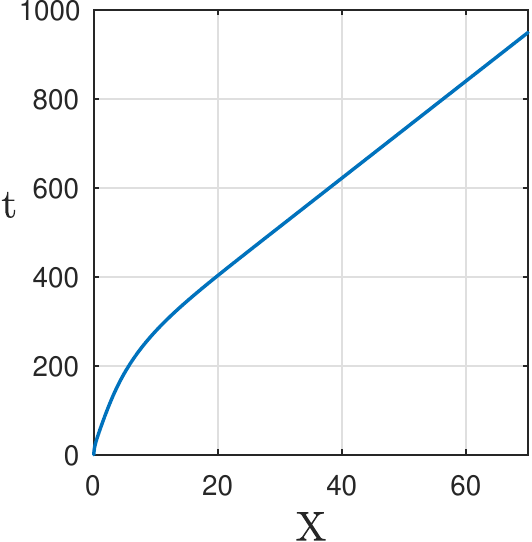}
	\caption{ Top: nontrapped soliton with the opposite polarity of the external force. Bottom (left): the space amplitude vs. crest position. Bottom (right): the crest position along the time. Parameters: $a=-0.5$, $w=10$ and $\epsilon=0.01$ and  $\Delta=-\frac{8}{15}\sqrt{|a|}$.}
	\label{Fig4}
\end{figure}
For the saddle point predicted by the dynamical system (\ref{DS}), we observe an interesting behavior where the depression soliton moves past the external force without changing its direction. Once it has passed the external force, the soliton amplitude stabilizes, and it continues to travel at a constant speed. These intriguing dynamics are clearly illustrated in Figure \ref{Fig4}. Furthermore, we can delve deeper into this phenomenon by examining the underlying physical mechanisms. When the depression soliton encounters the external force, the interaction between the soliton and the force causes a temporary disturbance in the soliton shape and amplitude. However, due to the nature of the external force and the characteristics of the soliton, once the soliton surpasses the external force, it reestablishes its original shape and amplitude, and continues its motion unaffected. This observation underscores the robustness and resilience of solitons in the presence of external perturbations.

In summary, the saddle point predicted by the dynamical system reveals an intriguing behavior where the depression soliton remains unaffected by the external force, allowing it to continue its travel with a constant speed and unchanged direction. These findings shed light on the remarkable stability and resilience of solitons in the presence of external perturbations.

%
%

\section{Conclusion}
In this paper, we have conducted a comprehensive investigation into the interaction between solitons and external forces within the framework of the Schamel equation. 
By incorporating an external force term into the Schamel equation, researchers can investigate how the presence of this force affects the dynamics and behavior of the ion acoustic waves. The external force could arise from various sources, such as electromagnetic fields, external potentials, or external driving mechanisms. Through a combined approach of asymptotic analysis and numerical simulations, we have gained valuable insights into the behavior of solitons under various scenarios.

Our findings reveal that the dynamics of solitons are highly dependent on the chosen parameters, specifically the soliton amplitude and crest position. We have observed that a soliton can exhibit different types of behavior: it can remain in a steady state if the parameters are appropriately selected, it can undergo oscillatory motion, bouncing back and forth near its initial position, or it can simply move away from its initial position without reversing its direction.

These results not only provide valuable qualitative insights into the soliton dynamics, but also demonstrate the effectiveness of the asymptotic theory in capturing the essential characteristics of soliton interactions with a broad range of external forces. Our analysis holds particularly qualitatively well when the length scale of the soliton is compared to that of the external force.

The implications of our study extend beyond the specific framework of the Schamel equation. The observed soliton behaviors and their dependence on parameters contribute to our understanding of nonlinear wave phenomena and offer practical applications in various fields, including optics, plasma physics, and fluid dynamics.

In conclusion, our combined asymptotic and numerical investigation sheds light on the rich dynamics of solitons interacting with external forces within the Schamel equation. By uncovering different types of soliton behavior and establishing the agreement between our findings and the asymptotic theory, we provide valuable insights into the broader understanding of soliton dynamics and their relevance in diverse scientific disciplines.

\section{Acknowledgements}
E.P. is thankful to the Russian Science Foundation grant 22-17-00153.. M.V.F is grateful to IMPA for hosting him as visitor during the 2023 Post-Doctoral Summer Program.

	\section*{Declarations}
	
	\subsection*{Conflict of interest}
	The authors state that there is no conflict of interest. 
	\subsection*{Data availability}
	
	Data sharing is not applicable to this article as all parameters used in the numerical experiments are informed in this paper.


\begin{thebibliography}{999}
	
	\bibitem{Ali:2017}{Ali R, Saha A, Chatterjee, P } 
	{Analytical electron acoustic solitary wave solution for the forced KdV
equation in superthermal plasmas.}
	{\it Plasma Phys.} {\bf 1972} 9(3), 122106. 



				\bibitem{Baines} {Baines, S.} 
      { Topographic effects in stratified flows.}
	{\it  Cambridge University Press, Cambridge.} {\bf 1995}. 
	
	\bibitem{Beiglbock:2000} {Beiglbock W,  Eckmann JP, Grosse H,  Loss M,  Smirnov S, Takhtajan L, Yngvason J.} 
      { Concepts and Results in Chaotic Dynamics.}
	{\it  Springer, Berlin.} {\bf 2000}. 
	
	\bibitem{Chowdhury:2018}{ Chowdhury S, Mandi L,  Chatterjee P } 
	{Effect of externally applied periodic force on ion acoustic waves
in superthermal plasmas.}
	{\it Phys. of Plasma.} {\bf 2018} 25, 042112. 	
	
				
	

	
		
			\bibitem{Ermakov}
	{Ermakov A.; Stepanyants, Y.} 
	Soliton interaction with external forcing within the Korteweg-de Vries equation
	{\it Chaos.}  {\bf 2019}, {29}, 013117.
	
		\bibitem{Marcelo-Paul-Andre}{Flamarion MV, Milewski PA,  Nachbin A.} 
				{Rotational waves generated by current-topography interaction.}
				{\it Stud Appl Math.} 2019;142:433-464. 
	
			\bibitem{Collisions}{Flamarion,  M.V.; Ribeiro-Jr, R.} 
		    {Solitary water wave interactions for the Forced Korteweg-de Vries equation.}
	{\it Comp. Appl. Math.} {\bf 2021}, 40, 312.
	
	
			\bibitem{Capillary2}{Flamarion, M.V.; Ribeiro-Jr, R.} 
        {Gravity-capillary flows over obstacles for the fifth-order forced Korteweg-de Vries equation.}
	{\it J. Eng. Math.} {\bf 2021}, 129, 1-17.

	
	
	
	\bibitem{COAM}{Flamarion, M.V.} 
		    {Generation of trapped depression solitary waves in gravity-capillary flows over an obstacle.}
	{\it Comp. Appl. Math.} {\bf 2022}, 41, 31.
	
		\bibitem{COAM2}{Flamarion, M.V.; Ribeiro-Jr, R.} 
		    {Trapped solitary-wave interaction for Euler equations with low-pressure region.}
	{\it Comp. Appl. Math.} {\bf 2021}, 40, 20.
	
			\bibitem{Chaos:FP}{Flamarion, M.V.; Pelinovsky E.} 
		    {Soliton interactions with an external forcing: the modified Korteweg-de Vries framework.}
	{\it Chaos, Solitons \& Fractals.} {\bf 2022}, 165, 112889.
	
				\bibitem{Whitham1}{Flamarion, M.V.} 
        {Waves generated by a submerged topography for the Whitham equation.}
	{\it Int. J. Appl. Comput. Math.} {\bf 2022}, 8, 257.
	
					\bibitem{Whitham2}{Flamarion, M.V.} 
        {Trapped waves generated by an accelerated moving disturbance for the
Whitham equation.}
	{\it Partial Differential Equations in Applied Mathematics.} {\bf 2022}, 5, 100356.
	

   \bibitem{Flamarion-Pelinovsky:2022a}{Flamarion MV, Pelinovsky E.} 
	{Solitary wave interactions with an external periodic force: The extended Korteweg-de Vries framework.}
	{\it Mathematics.} {\bf 2022}, 10, 4538.
	
	
	
	

	
	


	\bibitem{Grimshaw94}
	{Grimshaw, R.; Pelinovsky,  E.;  Tian, X.} 
	Interaction of a solitary wave with an external force.
	{\it Physica D.}  {\bf 1994}, 77, 405-433.
	

	
	\bibitem{Grimshaw96}
	{Grimshaw R.; Pelinovsky, E.; Pavel, S.} 
	Interaction of a solitary wave with an external force moving with variable speed.
	{\it Stud. Appl. Math.}  {\bf 1996}, {142}, 433-464.
	
			

	
			\bibitem{Grimshaw:1993}
	{Grimshaw, R.; Malomed, B.A.; Tian, X.} 
	Dynamics of a KdV soliton due to periodic forcing.
	{\it Phys. Lett. A.}  {\bf 1993}, {179}, 291-298.
			
	
		\bibitem{Grimshaw2002}
	{Grimshaw, R.; Pelinovsky, E.} (2002)
	Interaction of a solitary wave with an external force in the extended Korteweg-de Vries equation.
	{\it Int. J. Bifurcat. Chaos.}  {\bf 2002}, {12}(11), 2409-2419.
	
	
	
	
	

	

%

%
		      \bibitem{Grimshaw86}{Grimshaw R,  Smyth N.} 
{Resonant flow of a stratified fluid over topography in water of finite depth.}
	{\it J. Fluid Mech.} {\bf 1986}, 169, 235-276. 
%


			\bibitem{Kim} {Kim, H.; Choi, H.} 
      {A study of wave trapping between two obstacles in the forced Korteweg-de Vries equation.}
	{\it J. Eng. Math.} {\bf 2018}, 108, 197-208. 

	\bibitem{Lee} {Lee, S.} 
      {Dynamics of trapped solitary waves for the forced KdV equation.}
	{\it Symmetry.} {\bf 2018}, 10(5), 129. 

\bibitem{LeeWhang} {Lee, S.;  Whang, S.} 
      { Trapped supercritical waves for the forced KdV equation with two bumps.}
	{\it Appl. Math. Model.} {\bf 2015}, 39, 2649-2660. 
	
					\bibitem{Malomed:1993}{Malomed, B.A.} 
		    {Emission of radiation by a KdV soliton in a periodic forcing.}
	{\it Phys. Lett. A.} {\bf 1993}, 172, 373-377.
	
			\bibitem{Paul}{Milewski PA. } 
	{The Forced Korteweg-de Vries equation as a model for waves generated by topography.}
	{\it Cubo Math J.} {\bf 2004}, 6, 33-51. 

%
\bibitem{Mushtaq:2006}{ Mushtaq A, Shah HA.} 
	{Study of non-Maxwellian trapped electrons by using generalized (r,q) distribution function and their effects on the dynamics of ion acoustic solitary wave
.}
	{\it Phys. of Plasma.} {\bf 2006} 13, 012303. 	
	
	
		\bibitem{Nozaki:1983}{Nozaki K, Bekki N. } 
	{Chaos in a Perturbed Nonlinear Schr{\: o}dinger Equation.}
	{\it Phys. Rev. Lett.} {\bf 1983}, 50, 1226. 
	
	
%

	
		\bibitem{Pelinovsky:2002}
	{Pelinovsky E.} (2002)
	Autoresonance processes under interaction of solitary waves with
the external fields.
	{\it Int. J Fluid. Mech. Res.}  {\bf 2002}, {30}(5), 493-501.
	

\bibitem{Saha:2015a}{Saha A, Chatterjee P. } 
	{Qualitative structures of electron-acoustic waves in an unmagnetized plasma with q-nonextensive hot electrons.}
	{\it Eur. Phys. J. Plus.} {\bf 2015}, 130, 222. 	
	
	\bibitem{Saha:2015b}{Saha A, Chatterjee P. } 
	{Solitonic, periodic, quasiperiodic and chaotic structures of dust ion acoustic waves in nonextensive dusty plasmas.}
	{\it Eur. Phys. J. D.} {\bf 2015}, 69(9), 203. 	



\bibitem{Schamel:1972}{Schamel H. } 
	{Stationary solitary, snoidal and sinusoidal ion acoustic waves.}
	{\it Phys. of Plasma.} {\bf 2017}, 24, 377-387. 


\bibitem{Schamel:1973}{Schamel H. } 
	{A modified Korteweg-de Vries equation for ion acoustic wavess due to resonant electrons.}
	{\it Journal of Plasma Physics.} {\bf 1973}, 14, 905. 
	
		\bibitem{Trefethen:2000}{Trefethen, L.N.} 
        {\it Spectral Methods in MATLAB.}
	{Philadelphia: SIAM;} 2001.
	
	

	\bibitem{Williams:2014}{Williams G,   Verheest F,   Hellberg MA,  Anowar MGM,   Kourakis I} 
        {A Schamel equation for ion acoustic waves in superthermal plasmas.}
	{\it Phys. of Plasma.} {\bf 2014}, 21(9), 092103 . 
	
					\bibitem{Williams:1997} {Williams GP.} 
      { Chaos Theory Tamed .}
	{\it  Joseph Henry, Washington.} {\bf 1997}. 
	
		\bibitem{Wu1}{Wu TY.} 
        {Generation of upstream advancing solitons by moving disturbances.}
	{\it J Fluid Mech.} {\bf 1987}, 184, 75-99. 
	
	
%
%
%
	
	




		
%
%

         
	

	
	
%



	


%


	

		
		
		
	\end{thebibliography}
\end{document}